\documentclass[aps,prb,twocolumn,floatfix]{revtex4}
\begin{document}
\title{The breakdown of the mean-field  description
of the Nagaoka phase}

\author{Fabio L. Braghin}
\affiliation{International Center for Condensed Matter Physics,
Universidade de Brasilia, Caixa Postal 04513, 70904-970 Brasilia,
DF, Brazil}
\author{Alvaro Ferraz}
\affiliation{International Center for Condensed Matter Physics,
Universidade de Brasilia, Caixa Postal 04513, 70904-970 Brasilia,
DF, Brazil}
\author{Evgueny A. Kochetov}
\affiliation{Laboratory of Theoretical Physics, Joint Institute for
Nuclear Research, 141980 Dubna, Russia\\ International Center for
Condensed Matter Physics, Universidade de Brasilia, Caixa Postal
04513, 70904-970 Brasilia, DF, Brazil}

\begin{abstract}
We discuss the relevance of the improved mean-field slave-fermion theory
to describe the Nagaoka ($U=\infty$)
limit of the Hubbard model. In this theory
the crucial on-site constraint of no double electron occupancy
is taken into account rigorously prior to
the mean-field approximation.
At one-loop approximation the effective mean-field action shows a
long-range ferromagnetic order over the whole doping range.
This indicates that the slave-fermion mean-field theory
does not constitute an appropriate framework to describe the physics of the Nagaoka phase.
We discuss the drawbacks of this mean-field theory and present
some results on the derivation of a
 low-energy effective spin action to describe the Nagaoka phase
 beyond the mean-field approximation.
\end{abstract}
\pacs x 74.20.Mn, 74.20.-z
\maketitle

\section{Introduction}

It is now widely believed that the essential physics of strongly correlated lattice electrons is
encoded in the local constraint of no double electron occupancy. This constraint prohibits  double
electron occupancy of a lattice site due to the large local Coulomb repulsion between hopping electrons.
The on-site Hilbert space is thus restricted to states with at most one electron per site. Such a modification
of the underlying Hilbert space results in dramatic consequences for the low-energy properties of the relevant
electron systems revealing a rich and unusual physics in this limit.

Formally, given a local electron operator $c_{i\sigma}$ with the spin projection $\sigma=\uparrow,\downarrow$,
the local no double occupancy (NDO) constraint  reads
\begin{equation}
\sum_{\sigma}c^{\dagger}_{i\sigma}c_{i\sigma}\le 1.
\label{0.1}\end{equation}
Let us denote by $\tilde c_{i\sigma}$
the projected electron operators that satisfy this requirement.
Since the inequality (\ref{0.1}) is hard to deal with analytically,
one can try to
circumvent this difficulty by turning
Eq.(\ref{0.1}) into an equality at the expense of the introduction of some redundant degrees of freedom.
The two well-known ways to accomplish this are the so called slave-fermion (SF) and slave-boson (SB) representations
of the constraint electron operator. Namely, by introducing the "slave boson" \cite{sb}, one decomposes the on-site
constraint electron operator in the form,
\begin{equation}
\tilde c_{i\sigma}= b^{\dagger}_if_{i\sigma},
\label{0.2}\end{equation}
where $b_i$ is a charged spinless boson (holon), while $f_{i\sigma} $ ia a neutral, spin $1/2$ fermion operator (spinon)
satisfying the NDO constraint
\begin{equation}
 b^{\dagger}_ib_i +\sum_{\sigma}f^{\dagger}_{i\sigma}f_{i\sigma}=1.
\label{0.3}\end{equation}

Alternatively, one can also introduce a spinless fermion $f_i$
to describe the charge degree of freedom and a spinful
boson $b_{i\sigma}$ to keep track of the spin degree of freedom.
This is the "slave fermion" approach \cite{sf},
\begin{equation}
\tilde c_{i\sigma}= b_{i\sigma}f^{\dagger}_{i}.
\label{0.4}\end{equation}
The NDO constraint now reads
\begin{equation}
f^{\dagger}_if_i +\sum_{\sigma}b^{\dagger}_{i\sigma}b_{i\sigma}=1.
\label{0.5}\end{equation}
The electron spin operator,
$\overrightarrow{Q_{i}}=\frac{1}{2}%
\sum_{\sigma \sigma ^{\prime }}
\tilde c_{i\sigma}^{\dagger}\overrightarrow{\tau }%
_{\sigma \sigma ^{\prime }}\tilde c_{i\sigma ^{\prime }}$,
takes in this representation the form
\begin{equation}
\overrightarrow{Q_{i}}=\frac{1}{2}%
\sum_{\sigma \sigma ^{\prime }}b_{i\sigma}^{\dagger}\overrightarrow{\tau }%
_{\sigma \sigma ^{\prime }}b_{i\sigma ^{\prime }},
\label{Q1}\end{equation}
where Eq.(\ref{0.5}) has been used.
Here $\overrightarrow{%
\tau }^{\prime}$s are the Pauli matrices.

There is an apparent U(1) gauge redundancy in the decompositions of the projected electron operator as
given by Eqs.(\ref{0.2},\ref{0.4}). Namely, the local gauge transformation generated by the NDO constraint
\begin{equation}
f_i\to f_ie^{i\theta_i}, \, b_i\to b_ie^{i\theta_i}
\label{local}\end{equation}
keeps the representations (\ref{0.2}) and (\ref{0.4}) intact. The slave-particle gauge theory is infinitely
strong coupled since there is no kinetic energy for the bare gauge field. Consequently, one should exercise
some care in applying {\it an ad hoc}
mean-field (MF) approximation to treat the slave-particle theories. Such an approach  can only be justified
provided that the low-energy gauge coupling gets effectively renormalized to some finite and, potentially,
weak coupling. However, this seems hardly to be the case in the slave-particle theories \cite{nayak},
which in turn casts some doubts on the results obtained within the MF slave-particle theories. Indeed,
both the SF and SB theories should in principle produce physically identical results for the t-J model of
strongly correlated electrons, however they give in the MF approximation very different phase diagrams \cite{4,kf}.

The strong electron correlations are at work to
 full extent in the so called Nagaoka ($U=\infty$)
limit of the Hubbard model. Indeed, in this case an infinitely strong Coulomb repulsion strictly prohibits
the double electron occupancy of the lattice sites, and the NDO constraint becomes of the utmost importance.
In the infinite $U$ limit, the Hubbard Hamiltonian reduces to
\begin{equation}
H=-\sum_{ij,\sigma}t_{ij}\tilde c^{\dagger}_{i\sigma}\tilde c_{j\sigma}+\mu\sum_{i\sigma}(1-\tilde c^{\dagger}_{i\sigma}
\tilde c_{i\sigma}),
\label{1.1}\end{equation}
where $t_{ij}$ is a symmetric matrix whose elements represent the hopping amplitude $t$ between the
nearest-neighbor sites and are zero otherwise. We have also introduced the chemical potential $\mu$ to control
the total number of vacancies (holes).
Despite its seemingly simple form, this Hamiltonian cannot be diagonalized due to the fact that the
projected electron operators fulfill  complicated commutation relations
resulting from the explicit
manifestation of strong correlations.

Since the Coulomb repulsion $U$, whatever large
it may be,  is in practice always finite, the one band
$U=\infty$ Hubbard model is basically a toy model and does not describe any specific material\cite{optics}. Nevertheless, it captures an extreme limit for the physics of strong electron interactions. Had the projected electron operators in (\ref{1.1}) been replaced by their related conventional operators, the model would have been reduced to a system of noninteracting electrons revealing thereby a trivial physics. On the other hand, the physics behind the model (\ref{1.1}) is certainly far from  trivial.
Indeed, Nagaoka \cite{nagaoka} proved a theorem stating that for one hole
the ground state of the $U=\infty$ Hubbard model is a fully saturated
ferromagnet. This provides an interesting example of a quantum system
in which ferromagnetism appears as a purely kinetic energy effect with
 hole hoping
(itinerant ferromagnetism)
emerging as a result of the strong correlations from the NDO constraint.

Unfortunately, despite extensive work over many years, this model and
itinerant ferromagnetism are still poorly understood.
One of the important questions  to be addressed concerns the
thermodynamic stability of the Nagaoka phase.
That is, whether or not the Nagaoka state is stable when
the density of holes is finite in the thermodynamic limit.
There are arguments both for \cite{RY89,LONG-rasetti-94,richmond,36,21,44}
and against
\cite{suto,tian91,PLO92,tasaki,becca-sorella2001}
the thermodynamic stability of the
Nagaoka phase
and comparisons between various approaches have been made carefully (for a
recent example, see \cite{kotl}).
The basic problem that prevents one from reaching a definite conclusion
on that
is  the large-$U$ limit or, equivalently,
the local NDO constraint  which
is very difficult to deal with in a controlled way.
For example,
the numerical studies of this problem,
although well developed for the Hubbard model,
cannot incorporate the large-$U$ limit \cite{scalapino}.
Analytical approaches, on the other hand,
basically imply a MF treatment
in which the local NDO constraint is uncontrollably
replaced by a global condition.

In this sense, it is interesting to note, that there are
several works considering different
variational approaches
\cite{richmond,LONG-rasetti-94,shastry,vonderlinden-etal-91,basile-etal-90,hanisch-etal-97,becca-sorella2001}
in which the NDO constraint is automatically built
into the trial state vector.
In general, the following observation holds: variational estimations
that involve
more realistic refined trial wave-functions result in a smaller
value of the critical
hole concentration.

One may therefore think that a straightforward application
of the MF theory that treats
electron correlations on average
is not likely to
produce reliable results.
Specifically, in the present paper we show that the
treatment of the Nagaoka phase within the MF SF approximation
results in the qualitatively incorrect conclusions.

We outline below the main motivation of this work.
The standard
SF MF theory which treats the NDO
constraint only 
on average is known to predict a
 stable FM phase for the $U=\infty$ Hubbard model
over a finite though unphysically large doping range.
Since the local electron correlations encoded in the NDO constraint are surely of the utmost importance, if one attempts to deal with 
the properties of this model in a correct and reliable way, one
would naturally think that a proper treatment
of the constraint
prior to the MF approximation could improve that MF result shifting the critical hole concentration
towards a  much smaller and physically reasonable value.
 In this paper we show that this is not the case. On the
contrary, the improved MF theory predicts a
 stable FM phase over the whole doping range.

Two main routes which redirects 
of this result are discussed in Sections II and III. 
We argue that the finite critical hole concentration predicted
by the standard MF theory is just an artifact of
the uncontrolled treatment of the NDO constraint. 
Moreover, the theory based on the MF treatment of
the spin degrees of freedom affects the physics of the problem
in a qualitative way. Not only the NDO constraint needs to be fulfilled.
It is clearly very important to treat the electron
correlations encoded by that constraint in
a proper way considering both charge and spin degrees of
freedom on equal footing as dynamical rather than the MF variables.
Some preliminary results in this direction are reported in Section IV.

\section{MF SF treatment of the NDO constraint}

In this section we briefly go over the conventional MF SF
theory of the ferromagnetic phase
in the $U=\infty$ Hubbard model as given by, e.g.,
Boies {\it et al} \cite{canada}.

To set up the stage, let us start with the SF representation
of the projected electron
operators given by Eqs.(\ref{0.4},\ref{0.5}). The Hamiltonian (\ref{1.1})
can be then rewritten in the form
\begin{equation}
H=\sum_{ij,\sigma}t_{ij}f^{\dagger}_jb^{\dagger}_{i\sigma}b_{j\sigma}f_i+\mu\sum_{i}f^{\dagger}_if_i,
\label{2.1}\end{equation}
where Eq. (\ref{0.5}) has been used.
The partition function takes the form
\begin{equation}
Z=\int D\lambda D\bar\psi D\psi\prod_{\sigma} D\bar z_{\sigma}Dz_{\sigma}
e^{S(\psi,z, \lambda)},
\label{2.2}\end{equation}
with the action $S=\int_0^{\beta}d\tau{\cal L}_{SF}(\tau)$ and the Lagrangian
\begin{eqnarray}
{\cal L}_{SF}(\tau)&=&\sum_{ij}\bar{\psi}_j[(-\partial_{\tau}-\mu+\lambda_i)\delta_{ij}-t_{ij}\sum_{\sigma}\bar z_{i\sigma}z_{j\sigma}]\psi_i
\nonumber\\
&+&\sum_{i\sigma}\bar z_{i\sigma}(-\partial_{\tau}+\lambda_i)z_{i\sigma} -\sum_i\lambda_i.
\label{2.3}\end{eqnarray}
Here $z_{i\sigma}$ and $\psi_i$ stand for complex numbers and complex Grassmann parameters, respectively.
The purely imaginary field $\lambda_i$ has been introduced to enforce the local NDO constraint,
\begin{equation}
\bar\psi_i\psi_i+\sum_{\sigma}\bar z_{i\sigma}z_{i\sigma}=1.
\label{cnstr}\end{equation}
In this representation the electron spin operator (\ref{Q1}) becomes
\begin{equation}
\overrightarrow{Q_{i}}=\frac{1}{2}
\sum_{\sigma \sigma ^{\prime }}\bar z_{i\sigma}\overrightarrow{\tau }
_{\sigma \sigma ^{\prime }}z_{i\sigma ^{\prime }}.
\label{Q2}\end{equation}

\subsection{Zeroth-order approximation}

At the zeroth-order MF approximation
boson variables take a macroscopic value and all
fluctuations in space and imaginary time are neglected, i.e.
\begin{equation}
z_{i\sigma}(\tau)=z^{(0)}_{\sigma}, \quad \lambda_i(\tau)= \lambda^{(0)}.
\label{2.4}\end{equation}
Under this condition, the fermionic degrees of freedom in (\ref{2.2}) can easily be integrated out to yield
for the free energy ${\cal F}:=-\frac{1}{N\beta}\log Z,$
\begin{eqnarray}
{\cal F}_{SF}&=&-\frac{1}{N\beta}\sum_{\vec k}\log(1+e^{-\beta(E(\vec k)+\mu-\lambda^{(0)})})\nonumber\\
&-& \sum_{\sigma} |z^{(0)}_{\sigma}|^2\lambda^{(0)}+\lambda^{(0)}-\mu\delta,
\label{2.5}\end{eqnarray}
where $$E(\vec k)=(\sum_{\sigma} |z^{(0)}_{\sigma}|^2)t_{\vec k}$$ and $\delta$ is
the average density of holes.
The ground-state saddle point equations
determining $z_{\sigma}^{(0)},\, \mu$ and $\lambda^{(0)}$
describe a saturated ferromagnet, $m\propto (1-\delta)$, where the average magnetization is given by $m=\frac{1}{2}\sum_{\sigma}\sigma|z^{(0)}_{\sigma}|^2$.
Such a solution exists over the whole doping range, $0\le \delta < 1.$

It is interesting to note that due to the global MF constraint
$$\delta+\sum_{\sigma}\bar z^{(0)}_{\sigma}z^{(0)}_{\sigma}=1,$$
the spin variables $z^{(0)}_{\uparrow}$ and
$z^{(0)}_{\downarrow}$ cannot vanish
simultaneously.
This observation rules out the possibility of having,
within that approach,
a phase transition into a disordered phase
by identifying the set of macroscopic spin amplitudes,
$z^{(0)}_{\sigma}$, as the relevant order parameter.

\subsection{One-loop approximation}

The one-loop correction to the effective action can in principle
alter this result destabilizing the zeroth-order MF theory at some critical hole concentration, $\delta_c.$
In the MF SF theory the one-loop approximation takes the form \cite{canada}:
\begin{equation}
z_q^{\sigma}=z^{(0)}_{\sigma}\delta_{q,0}+\delta z_q^{\sigma},
\label{c1}\end{equation}
\begin{equation}
\lambda_q=\lambda^{(0)}\delta_{q,0}+\delta\lambda_q,
\label{c2}\end{equation}
where the zeroth-order MF solution,
$z^{(0)}_{\downarrow}=0$ and  $z^{(0)}_{\uparrow}=\sqrt{1-\delta}.$
Equation (\ref{c1}) tells us that quantum fluctuations of the
magnetic "order parameter"
are supposed to be small.
Besides that, Eq.(\ref{c2}) implies that the constraint-generated gauge field
is slowly fluctuating function around its MF constant value.
Assuming this,
one can substitute Eqs. (\ref{c1},\ref{c2}) into the
SF action to perform the fermion integral (\ref{2.2}) at one-loop
level.
In this way, one arrives at a purely bosonic effective action.
Analyzing then the stability criterion for the remaining bosonic integral over
$\delta z_{\sigma}$ and $\delta\lambda$,
one can evaluate
the critical value, $\delta_c\approx 0.7$, at which the zeroth-order MF theory becomes unstable
against quantum fluctuations.

However, this value of $\delta_c$ is too much large to agree with the recent results
of the variational Monte Carlo studies \cite{36} ($\delta_c=0.38$) and  small cluster calculations \cite{21}
($\delta_c=0.22$).
Note that the smallest value of $\delta_c$ obtained so far is $\delta_c=0.17$ \cite{44}.
This MF value also contradicts the exact result obtained in
infinite space dimension \cite{16}.

The one-loop MF result can be improved, in principle,
by imposing the crucial NDO constraint
prior to the MF calculations.
Surprisingly this is not the case.
On the contrary, as we show in the next section an
improved MF one-loop theory
predicts exactly the same result that follows directly
from the zeroth-order MF approximation:
the FM phase is stable in the whole doping range, $0\le \delta < 1.$
This apparently indicates that the MF theory based on the MF treatment of
the spin degrees of freedom affects the physics of the problem
in a qualitative way.
Within that approach,
the emergent
long-range ferromagnetic order appears to be just an artifact of the MF
approximation, rather than a dynamically generated effect.

\section{ Slave-fermion effective action with the explicitly
resolved NDO constraint}

Let us turn back to the exact representation (\ref{2.2}).
As a first step to proceed, we
rigorously resolve the constraint
$\bar\psi_i\psi_i+\sum_{\sigma}\bar z_{i\sigma}z_{i\sigma}=1$.
As a result, we arrive at the SF path-integral representation
of the partition function that explicitly incorporates the NDO constraint.

The constraint can be explicitly resolved by making the
identifications
\begin{eqnarray}
z_{i\uparrow }&=&\frac{e^{i\phi_i}}{\sqrt{1+\overline{z}_iz_i
+\overline{\xi}_i\xi_i}},\
z_{i\downarrow }=\frac{z_ie^{i\phi_i}}{\sqrt{1+\overline{z}_iz_i
+\overline{\xi}_i\xi_i}},\
\nonumber \\
\psi_i&=&%
\frac{\xi_i e^{i\phi_i}}{\sqrt{1+\overline{z}_iz_i+\overline{\xi}_i\xi}_i},
\label{3.1}
\end{eqnarray}
with the variables $z_i,\xi_i $ and $\phi_i$ being
free of any further  constraints.

Note that the local gauge transformation (\ref{local})
reasserts itself in the form,
$\phi_i\to \phi_i+\theta_i$. In contrast, the projected
variables
$$z_i=z_{i\downarrow}/z_{i\uparrow},\quad \xi_i=\psi_i/z_{i\uparrow}$$
are seen
to be manifestly gauge invariant.
As we already mentioned, this gauge symmetry
is a consequence of the redundancy in parametrizing the
electron operator in terms
of the auxiliary boson/fermion fields. The gauge ambiguity
related with the redundancy
of the SF representation is  now expressed by a single variable $\phi_i$.

The domain of the flat measure in (\ref{2.2}) that involves
the spin up bosonic fields can be rewritten
at every lattice site
as $D\bar z_{i\uparrow}Dz_{i\uparrow}=D|z_{i\uparrow}|^2D\phi_i$.
The $|z_{i\uparrow}|^2$ field can easily be integrated out from
Eq.(\ref{2.2}) thanks to the constraint (\ref{cnstr}).
Since the action in (\ref{2.2}) is U(1) gauge invariant and hence
independent of $\phi_i$,  the integration
over $\phi_i$ results merely in the appearance of some numerical factor
(volume of the gauge group) that can be taken care of
by a proper normalization of the partition function.
For the remaining integration we have
(the site dependence for the moment being
suppressed),
$$Dz_{\downarrow}D\bar z_{\downarrow}D\psi D\bar\psi= {\rm sdet}
\|\frac{\partial(z_{\downarrow},\bar z_{\downarrow},\psi,\bar\psi)}
{\partial(z, \bar z, \xi,\bar\xi)}\|\, d z d \bar z d\xi d \bar\xi.$$
The Jacobian of this transformation (superdeterminant) was evaluated
in~\cite{fkm} to yield
$${\rm sdet} \|\frac{\partial(a_{\downarrow},\bar a_{\downarrow},f,\bar f)}
{\partial(z,\bar z, \xi,\bar\xi)}\|=\frac{1}{1+|z|^2+{\bar\xi}\xi}.$$

Putting everything together we get a new representation for the SF
partition function (\ref{2.2})
with the local NDO constraint being explicitly resolved to give
\begin{equation}
Z=\int D\mu (z,\xi )\
e^{S(z, \xi)}, \label{3.2}
\end{equation}%
where
\begin{equation}
D\mu (z,\xi )=\prod_{j,t}\frac{d\bar z_j(t)dz_j(t)}{2\pi i}\,
\frac{d\bar\xi_j(t)d\xi_j(t)}{1+|z_j|^2+\bar\xi_j\xi_j}
\label{3.2.2} \end{equation} stands for
the measure with the boundary conditions,
$z_j(0)=z_j(\beta),\, \xi_j(0)= -\xi_j(\beta).$ The action now reads
\begin{eqnarray}
S(z,\xi)&=&\frac{1}{2}\sum_j\int_0^{\beta}\frac{\dot{\bar z_j}z_j-\bar
z_j\dot z_j
+\dot{\bar\xi_j}\xi_j-\bar\xi_j\dot\xi_j}{1+|z_j|^2+\bar\xi_j\xi_j}dt \nonumber \\
&&-
\int_0^{\beta}H^{cl}dt. \label{3.3}\end{eqnarray}
The first
part of the action (\ref{3.3}) is a purely kinematic term that
reflects the geometry of the underlying phase space while the
classical Hamiltonian  becomes,
\begin{eqnarray}
H^{cl}
=t\sum_{ij}\frac{\bar\xi_j\xi_i(1+z_j
\bar z_i) + H.c.}
{(1+|z_i|^2+\bar\xi_i\xi_i)(1+|z_j|^2+\bar\xi_j\xi_j)}
\label{3.4}\end{eqnarray}
The new set of the gauge invariant variables, $(z,\xi)$, explicitly resolve
the NDO constraint at the apparent expense of a more
complicated compact phase space for the projected
electron operators.

Finally, we make a change of variables to decompose the full measure given by (\ref{3.2.2})
into the product of the conventional spin and fermion measures,
\begin{eqnarray}
D\mu _{spin}(\bar z,z)&=&\prod_{j,t}\frac{d\bar z_j(t)dz_j(t)}
{2\pi i(1+|z_j(t)|^2)^2},\nonumber \\
D\mu _{fermion}(\bar\xi,\xi)&=&\prod_{j,t}
d\bar\xi_j(t)d\xi_j(t), \nonumber
\end{eqnarray}
respectively.
Such a reparametrization can be taken to be
\begin{equation}
z_i\rightarrow z_i,\ \xi_i \rightarrow \xi_i \sqrt{1+|z_i|^{2}}.  \label{3.5}
\end{equation}
Up to an inessential factor which redefines the chemical potential, we get
\begin{equation}
D\mu \rightarrow D\mu_{spin}(\overline{z},z)\times D\mu_{fermion}(%
\overline{\xi },\xi ),  \label{3.6}
\end{equation}
and the effective action becomes
\begin{eqnarray}
S&=&\sum_i\int_0^{\beta}ia_i(\tau)d\tau-\sum_i\int_0^{\beta}\bar\xi_i
\left(\partial_{\tau}+\mu+ia_i\right)\xi_id\tau \nonumber\\&-&
\int_0^{\beta}H^{cl}d\tau 
\label{3.66}\end{eqnarray}
This action involves
the U(1)-valued connection one-form of the magnetic monopole
bundle that can formally be interpreted as a spin
"kinetic" term,
$$ia=-\langle z|\partial_t|z\rangle=\frac{1}{2}\frac{\dot{\bar z}z-\bar z\dot
z}{1+|z|^2},$$ with $|z\rangle$ being the su(2) coherent state (see Appendix).
This term is also frequently referred to as the Berry connection.
The dynamical part of the action takes the form
\begin{eqnarray}
H^{cl}=
t\sum_{ij}(\overline{\xi }_{j}\xi _{i}\langle
z_{i}|z_{j}\rangle +H.c.). 
\label{3.7}
\end{eqnarray}
Here $\langle z_{i}|z_{j}\rangle$ stands for an inner product of the su(2)
coherent states, $$\langle z_{i}|z_{j}\rangle =\frac{1+\overline{z}%
_{i}z_{j}}{\sqrt{(1+|z_{j}|^{2})(1+|z_{i}|^{2})}}.$$
The classical image of the on-site electron spin opertor (\ref{Q2}) reduces to
$$\vec Q^{cl}_i=\vec S_i^{cl}(1-\bar\xi_i\xi_i),$$
where $\vec S^{cl}$ is given by Eqs.(\ref{1a.3}).

The action (\ref{3.66}) is invariant under the global SU(2) rotations
that now take the form
\begin{eqnarray} z_i\rightarrow
\frac{uz_i+v}{-\overline{v}z_i+\overline{u}}, \quad
\xi_i\rightarrow
e^{i\zeta_i }\xi_i,\quad a_i\to a_i-d\zeta_i,
 \label{3.8}
\end{eqnarray}
where
\begin{eqnarray}
\zeta_i=-i\log \sqrt{\frac{-v\overline{z}_i+u}{-\overline{v}z_i+\overline{%
u}}}, \quad \left(\begin{array}{ll}
u & v \\
-\overline{v} & \overline{u}%
\end{array}
\right) \in \mathrm{SU(2)}.
\end{eqnarray}

Equations (25,26) provide a rigorous representation 
of the effective action of the $U=\infty$ Hubbard model.
One can therefore 
wonder if these equations lead to the expected result if one nullifies
 the $t_{ij}$ hoppings. 
In this limit the infinite $U$ Hubbard
model  (8) reduces to the exactly solvable reduced Hamiltonian:
$$H=\mu\sum_{i\sigma}(1-\tilde c^{\dagger}_{i\sigma}
\tilde c_{i\sigma})=\sum_{i}X^{00}_i,$$
where the Hubbard operator $X^{00}$ is represented by the
diagonal $3\times 3$ matrix with
 eigenvalues $0,0$ and $1$.  As a result, the partition function reduces simply
to:
$$Z=2+e^{-\beta\mu}.$$
As a check of the validity of our representation our equations 
(\ref{3.66},\ref{3.7})
should also recover this result.

To see that this is indeed the case, let us consider the on-site action (25) at $t=0$,
$$S=\int_0^{\beta}ia(\tau)d\tau-\int_0^{\beta}\bar\xi
\left(\partial_{\tau}+\mu+ia\right)\xi d\tau.$$
By a gauge transformation the potential $a(\tau)$ can be brought into a time independent form,
$$a\to a-\dot\phi=\frac{1}{\beta}\int_0^{\beta}ad\tau,$$ where
$$\phi(\tau)=-\frac{\tau}{\beta}\int_0^{\beta}ads +\int_0^{\tau}ads.$$ Note that $\phi(0)=\phi(\beta).$
The effective action then becomes
$$S=\int_0^{\beta}ia(\tau)d\tau-\int_0^{\beta}\bar\xi
\left(\partial_{\tau}+\bar{\mu} \right)\xi d\tau,$$
where $\bar{\mu}=\mu+\frac{1}{\beta}\int_0^{\beta}iad\tau.$ The partition function is now given by the path integral
$$Z=\int D\mu e^{S},$$ where the measure factor is represented by Eq.(24). Integrating out fermions yields
$$Z=\int D\mu_{spin} e^{\int_0^{\beta}ia(\tau)d\tau}(1+e^{-\beta\bar{\mu}}).$$ Since
$$\int D\mu_{spin} e^{\int_0^{\beta}ia(\tau)d\tau}=Tr_{spin}\,\hat I=2,$$ and $\int D\mu_{spin}=1,$ one finally gets
$Z=2+e^{-\beta\mu},$
which is indeed what it should be desired.

\subsection{Improved MF theory: zeroth-order approximation}

In this approximation
the bosonic spin variable $z_i(t)$ takes on a macroscopic value $z^{(0)}$.
Since $\langle z^{(0)}|z^{(0)}\rangle=1$, the spinless fermion MF Hamiltonian
(\ref{3.7}) reduces to
the representation
\begin{eqnarray}
H^{cl}=
t\sum_{ij}(\overline{\xi }_{j}\xi _{i}+H.c.). \label{t1}
\end{eqnarray}
This zeroth-order Hamiltonian does not depend on $z^{(0)}$ whose value merely determines a
direction of the total electron
magnetic moment,
\begin{equation}
\vec Q_{MF}=\vec S_{cl}(\bar z^{(0)},z^{(0)})(1-\delta),
\label{q}\end{equation}
with the explicit representation of the spin moment $S_{cl}(\bar z,z)$
being given in Appendix.

>From now on we take
the total electron magnetic moment
aligned along the $z$-axis. To achieve this, one needs to set
$z^{(0)}=0$ (see Appendix).
The Hamiltonian (\ref{t1}) along with Eq.(\ref{q})
describes a fully polarized ferromagnet for any hole concentration
$\delta<1.$ Therefore, the MF theory that takes care of the NDO constraint
at the outset, produces naturally,
at zeroth-order approximation, nearly
the very same result that is produced by the
zeroth-order MF theory (\ref{2.4})
which only treats the NDO constraint
globally.
Note, however, that in the representation (\ref{t1}), in contrast to
Eq.(\ref{2.5}), there is no renormalization
of the fermionic bandwidth.
Again, one  sees that
due to the NDO constraint, the MF treatment of the spin
dynamics, $z_i(t)=z^{(0)}$, automatically drives the system into an ordered
phase. Regardless of any specific value of the
constraint-free spin variable $z^{(0)}$, the system always stays
in the ordered ferromagnetic phase
as dictated by Eq. (\ref{q}).
The "order parameter" $z^{(0)}$, once again,
fails to  describe a phase transition out of the ferromagnetic phase.

\subsection{Improved MF theory: one-loop approximation}

In the present subsection we derive the one loop
approximation for the effective MF spin action,
with the local NDO  constraint built
in from the outset.

To this end we take into account the Gaussian fluctuatins of the bosonic spin variables
around their MF value,
\begin{equation}
z_i(t)=z^{(0)}+\delta z_i(t)=\delta z_i(t),
\label{z}\end{equation}
and expand the action
(\ref{3.66}) up to the quadratic order in the new $\delta z_i$ variables.
The fermionic path-integral can then be evaluated
to this order explicitly and we end up with the following
one-loop spin MF effective action,
\begin{eqnarray}
S&=& S_0+\Delta S,\label{t2}\end{eqnarray}
where $S_0$ represents the zeroth-order MF action, whereas
\begin{eqnarray}
\Delta S&=&
Tr \,\delta (ia)+ Tr G_{(t)}\delta\Sigma.
\label{t3}\end{eqnarray}
Here
$$\delta (ia_i)=\frac{1}{2}(\dot{\bar{\delta z_i}}\delta z_i-\bar{\delta z_i}\delta{\dot z}_i)$$ is a linearized
spin kinetic term,
$$\delta \Sigma_{ij}=-t_{ij}(\delta\bar z_j\delta z_i-\frac{1}{2}|\delta z_i|^2- \frac{1}{2}|\delta z_j|^2),$$
and
$$G_{(t)}^{-1}=(\partial_t-\mu-t_{ij})$$ is a zeroth-order MF Green function.
The trace has to be carried out in both space and time indices.

In the momentum space the action (\ref{t3}) reads
\begin{eqnarray}
\Delta S&=&
\frac{1}{2}\sum_{\vec q}\int_0^{\beta}
(\dot{\bar{\delta z_{\vec q}}}\delta z_{\vec q}-\bar{\delta z_{\vec q}}
\dot\delta z_{\vec q})d\tau\nonumber\\
&-&\sum_{\vec q}\int_0^{\beta}\omega_{\vec q}\bar{\delta z_{\vec q}}
\delta z_{\vec q}d\tau,
\label{t4}\end{eqnarray}
where
\begin{eqnarray}
\omega_{\vec q}=\frac{1}{N}\sum_{\vec p}f_{\vec p}(
t_{\vec p+\vec q}-t_{\vec p}),
\label{t5}\end{eqnarray}
and $f_{\vec p}=(e^{\beta(t_{\vec p}+\mu)}+1)^{-1}$ is the Fermi
distribution function.
The action (\ref{t4}) corresponds to the bosonic spin-wave Hamiltonian,
\begin{eqnarray}
H=\sum_{\vec q}\omega_{\vec q}b^{\dagger}_{\vec q}b_{\vec q},
\quad [b^{\dagger}_{\vec q'},b_{\vec q}]
=\delta_{\vec q',\vec q}.
\label{t6}\end{eqnarray}
with the ferromagnetic spin-wave dispersion relation, $\omega_{\vec q}\propto \vec q^2,\, \vec q\to 0.$

These quantum fluctuations  cannot destabilize the zeroth-oreder MF solution,
since the excitation spectrum $\omega_{\vec q}$ is
a non-negative
function of the hole concentration $\delta$,
provided
$t_{\vec p}=t_{-\vec p}$. This can be proven in the following way: \cite{marcin}
\begin{eqnarray}
\omega_{\vec q}&=&\frac{1}{N}\sum_{\vec p\in BZ}f_{\vec p}
(t_{\vec q+\vec p}-t_{\vec p}) \nonumber \\
&=&  \frac{1}{2} \frac{1}{N}\sum_{\vec p\in BZ}f_{\vec p}
(t_{\vec q+\vec p}-t_{\vec p})+
\frac{1}{2}\frac{1}{N}\sum_{\vec p\in BZ}f_{\vec p}
(t_{\vec q+\vec p}-t_{\vec p}) \nonumber \\
\end{eqnarray}
In the second summation we substitute $\vec p \rightarrow - \vec p - \vec q   $.
In this way we get
\begin{eqnarray}
\omega_{\vec q}
&=&  \frac{1}{2} \frac{1}{N}\sum_{\vec p\in BZ}f_{\vec p}
(t_{\vec q+\vec p}-t_{\vec p})\nonumber\\
&+&
\frac{1}{2}\frac{1}{N}\sum_{\vec p\in BZ}f_{- \vec p -\vec q}
(t_{\vec q-\vec p -\vec q}-t_{-\vec p -\vec q}) \nonumber \\
&=&  \frac{1}{2} \frac{1}{N}\sum_{\vec p\in BZ}f_{\vec p}
(t_{\vec q+\vec p}-t_{\vec p})\nonumber\\
&+&
\frac{1}{2}\frac{1}{N}\sum_{\vec p\in BZ}f_{\vec p +\vec q}
(t_{\vec p}-t_{\vec p +\vec q}) \nonumber \\
&=&  \frac{1}{2} \frac{1}{N}\sum_{\vec p\in BZ}(f_{\vec p}-f_{\vec p +\vec q})
(t_{\vec q+\vec p}-t_{\vec p}) \nonumber
\end{eqnarray}
Since $f_{\vec p}$ is monotonically decreasing function of $t_{\vec p}$,
the quantity $(f_{\vec p}-f_{\vec p +\vec q})
(t_{\vec q+\vec p}-t_{\vec p}) $ is always non-negative.

In fact, the spin-wave stability condition,
$\omega_{\vec q}> 0,\, \vec q\ne 0$, is only a necessary condition
for the saturation of ferromagnetism \cite{mattis}.
However, it is possible to show that if the semiclassical analysis
is assumed to be true, this also becomes a sufficient condition.
Therefore, if
the MF (semiclassical) decomposition of the spin amplitude (\ref{z})
is taken for grant, the ensuing SF MF
theory is spin-wave stable over the whole doping range.

There is, however, a rigorous result that states that for a
 large enough
hole concentration the Nagaoka state posesses an instability \cite{shastry}.
This result is heavily based on
the consideration that
the spin degrees of freedom in the $U=\infty$ limit are in fact dynamical
variables rather than nearly
frozen spins in the background. As was
discussed above, the MF treatment of the spin degrees of freedom
automatically drives the system into the ordered FM phase.
Although in this case the SF MF theory works well,
it is applied to a substantially altered Hamiltonian that does not
seem to bear much in common with the original problem.

\section{One-loop approximation beyond MF theory}

In the preceeding section we derived the improved one-loop MF
theory to treat the Nagaoka's phase.
In the improved theory we imposed the
NDO constraint rigorously from the beginning
and, only after that, the MF approximation was applied to describe
the dynamics of the spin degres
of freedom.
The improved theory predicts a stable FM phase over the whole doping range. At the same time, the standard
SF MF theory that treats the NDO
constraint globally at the zeroth-order and locally at the one-loop order tells us that the FM phase is
stable over a finite though quite large doping range.
Since the improved MF theory should in any event provide a better description, one may conclude
that the predictions of the
standard MF theory are not reliable and are just an artifact of the
uncontrolled MF treatment.

In this sense, it is interesting to note, that in an earlier  paper \cite{richmond}
a variational principle was formulated
in such a way that
the NDO constraint was automatically built into the trial state vector
which considers that all the electron spins except one are aligned.
The remaining electron is kept frozen.
It can then rigorously be shown that at $U=\infty$
the Nagaoka ferromagnet is always stable for any $\delta$.
This qualitatively agrees with the result
following from the improved one-loop MF theory.

However, if the remaining electron is also allowed to hop around,
the FM ground state immediately becomes unstable for
large enough doping \cite{shastry}. All this indicates, that even
the improved MF approximation is too restrictive in the sense that
it qualitatively affects the physics
described by the model by unnecessarily freezing the spin degrees of freedom.
This leads to the physically
incorrect conclusion of predicting the FM phase for the whole
doping range \cite{css}.
It is evident that not only the NDO constraint needs to be fulfilled.
It is also very important to treat the electron
correlations encoded by that constraint in
a proper way considering both charge and spin degrees of
freedom on equal footing as dynamical variables.
It would therefore be very appealing to address
the problem of the thermodynamic properties of the Nagaoka phase starting
right from the low-energy effective quantum spin Hamiltonian rather than
from the quasi-classical MF spin
effective action. Technically
this approach is, however, quite complicated. In view of that we report in this section only some preliminary results.

Specifically, we derive the contribution to the low-energy effective spin action
of the $U=\infty$ Hubbard model up to the lowest nonvanishing order in the spin self-energy
$\Sigma_{ij}=-t\langle z_j|z_i\rangle$
beyond the MF approximation.
To this end, we rewrite action
(\ref{3.66}) in the form
\begin{eqnarray}
S&=& i\sum_{i}\int_0^{\beta}a_i(\tau)d\tau \nonumber\\
&+&\sum_{ij}\int_0^{\beta}\bar\xi_i(\tau)
G^{-1}_{ij}(\tau,\sigma)\xi_j(\sigma)d\tau d\sigma,\label{3.9}\end{eqnarray}
where
$$G^{-1}_{ij}(\tau,\sigma)=G^{-1}_{(0)ij}(\tau,\sigma)-ia_i(\tau)\delta_{ij}\delta(\tau-\sigma)+\Sigma_{ij}(\tau)
\delta(\tau-\sigma),$$
with $ \Sigma_{ij}= -t_{ij}\langle z_j|z_i\rangle$ and
$$G^{-1}_{(0)ij}(\tau,\sigma)=\delta_{ij}(-\partial_{\tau}-\mu)\delta(\tau-\sigma).$$

The fermionic degrees of freedom can formally be
integrated out to yield
\begin{eqnarray}
&&\int D\bar\xi
D\xi\exp{\left(\sum_{ij}\int_0^{\beta}\bar\xi_i(\tau)
G^{-1}_{ij}(\tau,\sigma)\xi_j(\sigma)dtds\right)}\nonumber\\
&=&\exp{Tr\log G^{-1}}\nonumber\\
&=&\exp{\left(Tr\log
G^{-1}_{(0)}+Tr\log(1-G_{(0)}ia+G_{(0)}\Sigma)\right)}.
\label{3.10}\end{eqnarray}
Here the trace has to be carried out
over both space and time indices.
Calculating explicitly
the contribution coming from the zero order Green's function, it yields:
$$Z_0:=Z_{t=a=0}=\exp{(Tr\log G^{-1}_{(0)})}$$
$$=\left(2\cosh\frac{\mu\beta}{2}\,e^{-\frac{\mu\beta}{2}}\right)^N,$$
which reproduces the exact result for the partition function
of N noninteracting spinless fermions,
$$Z_0=tr\,
e^{-\mu\int^{\beta}_0 \sum_if^{\dagger}_if_i}.$$

We now evaluate the contributions of the self-energy $\Sigma_{ij}$ and
of the
gauge potential $a_0(i)$
up to the first nonvanishing order.
This can be done in the usual way
by making a loop expansion  in the trace \cite{schakel,ranninger}.
Here we are interested in the lowest order
contribution that survives in the low-energy
and long-wavelength limit.
This limit consists in expanding the one-loop contribution
up to first order in $\partial_{\tau}$ and up to the
second order in $\vec R_i -\vec R_j$
implying that, eventually, we will set
$i\to j$.
This amounts to the so-called gradient expansion
corresponding to the low-energy and long-wavelength limit of the
action.
We obtain in this way
\begin{eqnarray}
&&Tr\log(1-G_{(0)}ia+G_{(0)}\Sigma)=\nonumber\\
&&-Tr(G_{(0)}ia)-Tr [\frac{1}{2}(G_{(0)}\Sigma
G_{(0)}\Sigma) +{\cal O}(\Sigma ^3)].\label{3.11}\end{eqnarray} Note that
$Tr(G_{(0)}iaG_{(0)}\Sigma)=0$ since $\Sigma_{ii}=0$.
Note that Eq.(\ref{3.11}) is invariant
under global rotations.
This immediately follows from the transformation
law of the  given by Eq.(\ref{3.8})
accompanied by the similar transformations from the
$\Sigma_{ij}$'s: $$\Sigma_{ij} \to e^{-i\zeta_j+i\zeta_i}\,\Sigma_{ij}.$$

The $a$-dependent term in Eq.(\ref{3.11}) contributes to the action
in the following way
$$-Tr(G_{(0)}ia)=-i\sum_{i}G_{(0)i}(0^{-})\int_0^{\beta}a_i(\tau)d\tau,$$
 where
$G_{(0)i}(0^{-}):=\lim_{\epsilon \to 0}G_{(0)i}(-\epsilon), \,\,
\epsilon > 0$ and
\begin{equation}
G_{(0)i}(\tau)=\frac{e^{-\mu\tau}}{1+e^{\mu\beta}}-\theta(\tau)e^{-\mu\tau}.
\label{3.12}
\end{equation}
The explicit representation
(\ref{3.12}) tells us that
\begin{equation}
Tr(G_{(0)}ia)=
{\cal O}(e^{-\mu\beta}),\quad \mu\beta >>1. \label{3.13}\end{equation}

Let us now turn to the second term in Eq.(\ref{3.11}). We get
$$-\frac{1}{2}Tr(G_{(0)}\Sigma G_{(0)}\Sigma)$$
$$=-\frac{1}{2}\sum_{ij}\!\!\int G_{(0)i}(t_1-t_2)\Sigma_{ij}(t_2)
G_{(0)j}(t_2-t_1)\Sigma_{ji}(t_1)dt_1dt_2.$$
Introducing new
variables, $\tau=\frac{t_1-t_2}{2}, \, \eta=\frac{t_1+t_2}{2}$,
and expanding the product
$\Sigma_{ij}(\eta +\tau)\Sigma_{ji}(\eta
-\tau)=\Sigma_{ij}(\eta)\Sigma_{ji}(\eta)+{\cal O}(\tau)$
(this
corresponds to the gradient expansion in imaginary time
\cite{ranninger}), this, to lowest order, reduces to
\begin{equation}
-\frac{1}{2}\sum_{ij}\int^{\beta}_{-\beta}G_{(0)}(\tau)G_{(0)}(-\tau)d\tau
\!\!\int^{\beta}_0\Sigma_{ij}(\eta)\Sigma_{ji}(\eta)d\eta
\label{2.11}\end{equation}
With the help of Eq.(\ref{3.12}) we get
\begin{equation}
-\frac{1}{2}\int^{\beta}_{-\beta}G_{(0)}(\tau)G_{(0)}(-\tau)d\tau
=\frac{\beta}{4}\frac{1}{\cosh^2(\beta\mu/2)}.\label{3.133}\end{equation}
Note that the energy scale is set by the chemical potential $\mu$, so that
the low-energy limit takes the form $\mu\beta\gg 1.$

The effective spin action is then given by the sum of all the terms
evaluated above:
\begin{equation}
Z_{eff}/Z_0=\int D\mu (z,\bar z)\ e^{S_{eff}},
\label{3.14}
\end{equation}
where the SU(2) invariant measure is
$$D\mu (z,\bar z)=\prod_{i,t}\frac{d\bar z_i(t)dz_i(t)}{2\pi
i(1+|z_i|^2)^2}$$
and the effective action:
\begin{eqnarray}
S_{eff}&=&i\sum_{i}\int_0^{\beta}a_i(\tau)\nonumber\\
&+&\sum_{ij}\int_0^{\beta}\frac{J_{ij}^{eff}}{2}|\langle
z_i|z_j\rangle|^2,\label{3.15}\end{eqnarray}
with the
long-wavelength limit $(i\to j)$ being implicit
throughout our calculation.
This action describes the
SU(2) invariant ferromagnetic Heisenberg model with the effective coupling
(see Appendix)
$$-J^{eff}_{ij}=-(\beta|t_{ij}|^2)/(2\cosh^2\frac{\beta\mu}{2})\le 0.$$
Since
\begin{equation}
J^{eff}_{ij}={\cal O}(e^{-\mu\beta}),\quad \mu\beta >>1
\label{r}\end{equation}
this result tells us that
the lowest order contribution to the free energy in $\Sigma$
shows no magnetic ordering,
provided $\mu\beta\gg 1.$
However, one can safely truncate the expansion of the effective action (40)
at second order, provided
$|t|\beta\ll 1$. This, however, does not generate a high-temperature expansion of the free energy.
The point being that the temperature is supposed to be "high" compared to the overall energy scale factor $t$.
However, it still might be low compared with the "intrinsic" energy scale which is set by $\mu$.
Physically, the limit $\mu\beta\gg 1$ corresponds to a
very small hole concentration\cite{zhang}.
Therefore,
our result (\ref{r}) provides us with a limited piece of information concerning the thermodynamic instability of the Nagaoka phase at finite temperature. It merely indicates that the FM order
exponentially decays away from half filling at any finite temperature $T=1/\beta\gg |t|$.

To address the issue of the thermodynamic instability of the Nagaoka phase at any $T\ge 0$
one should go back to the full series in
the  low-energy long-wavelength  expansion of the fermionic determinant
(\ref{3.11}) and analyse further the dynamics of the spin variables as well.
On the bipartite lattice with the nn interaction,
that series can be summed up to yield
the following contribution to the low-energy effective spin action:
\begin{eqnarray}
\Delta S=\frac{e^{-\mu\beta}}{2\beta}\int_0^{\beta}
dt \,Tr\,\cosh(2\beta\Sigma(t)), \quad \mu\beta>>1.
\label{3.16}\end{eqnarray}
The next step is now to calculate the long-wavelength asymptotic of this representation,
which is not a trivial
follow up and it still is in progress.
Note only that this result is
strongly dependent on the space dimensionality.
However, there exists a universal feature of the $U=\infty$
Hubbard model phase diagram that naturally emerges from Eq.(\ref{3.16}): the
paramagnetic state is thermodynamically stable at any finite temperature at $\delta =0$.
This can be derived by considering the limit $\mu\to\infty$ in Eq.(\ref{3.16}) at finite
values of $\beta$ and $t$.
This result agrees with the qualitative arguments presented in \cite{kotl}.
In contrast with that,
the MF treatment continue to predict in this case a fully polarized FM state.

\section{Conclusion}

In this paper, we formulate the improved SF MF theory to describe the FM phase of the Nagaoka
limit of the Hubbard model. It is clear that the physics behind the Nagaoka phase is
controlled by strong electron correlations. Those correlations are in turn encoded into the NDO
constraint. We improve the standard MF approach by taking
the NDO constraint rigorously into
consideration
 prior to the MF approximation. Once this is done we integrate out the fermionic
degrees of freedom under the condition that the spin degrees of freedom are considered at the MF level.
In this way we arrive at the one-loop MF theory of the Nagaoka phase that predicts a FM ordering over the
whole doping range. This result agrees with earlier variational estimates
which take the NDO constraint fully into account but
leaves some of the spin degrees of freedom completely frozen.
At the same time, the conventional SF MF theory,
that treats the constraint at the MF level, predicts the FM phase
over a finite though quite large doping range.
This result of the standard
MF theory thus appears just as
an artifact of the uncontrolled treatment of the NDO constraint.

We show that the SF MF theory automatically drives the system
into the FM ordered phase, and hence it cannot
describe the underlying physics even in a qualitatively correct way.
This happens because of the fact that the
spin degrees of freedom are only considered quasi-classically. However,
it is the quantum spin dynamics that ultimately determines the magnetic properties of the Nagaoka phase.
Therefore, to get any physically reasonable result one should
enforce both the NDO constraint rigorously and treat the
dynamics of the spin degrees of freedom beyond the MF approximation.

As a first step in this direction, we derived the truncated low-energy long-wavelength
effective spin action with the spin variables being treated beyond the MF approximation.
This approximation corresponds to the expansion of the effective free energy in powers of a small parameter
$|t|\beta$ close to half filling.
It turns out that
treating the spin dynamics beyond the MF approximation
completely destroys the FM order predicted by the SF MF theory.
Despite of that, this result does not prove
the thermodynamic instability of the Nagaoka phase. To address this issue one should go back
to the full non-truncated
low-energy long-wavelength effective spin action as discussed in the previous sections
of the paper.
Nevertheless, we can add with certainty that
the SF MF approach produces spurious results and is therefore not reliable
for the description of the Nagaoka ferromagnetism.

\section{Acknowledgments}

This work was partially supported by the
Brazilian Ministry of Science and Technology and by CNPq.

\section{Appendix: su(2) algebra and coherent states}

Consider the su(2) algebra in the lowest $s=1/2$ representation:
\begin{equation}
[S_z,S_{\pm}]=\pm S_{\pm},\quad [S_{+},S_{-}]=2S_z, \quad \vec
S^2=3/4.\label{1a.1}\end{equation} Acting with the ``lowering``
spin operator $S^{-}$ on the ``highest weight`` state
$|\uparrow\rangle$ we get the normalized su(2) CS parametrized by
a complex number $z$
\begin{equation}
|z\rangle=\frac{1}{\sqrt{1+|z|^2}}\exp( zS^{-})|\uparrow\rangle=
\frac{1}{\sqrt{1+|z|^2}}(|\uparrow\rangle +z|\downarrow\rangle).
\label{1a.2}\end{equation} In the basis spanned by the vectors
$|\uparrow\rangle, \,|\downarrow\rangle$ we have
$S_{+}=|\uparrow\rangle|\langle\downarrow|,
\,S_{-}=|\downarrow\rangle|\langle\uparrow|, \,
S_z=\frac{1}{2}(|\uparrow\rangle|\langle\uparrow|-|\downarrow\rangle|\langle\downarrow|).$
The CS symbols of the su(2) generators are then easily evaluated
to be ($S^{cl}:=\langle z|S|z\rangle$):
\begin{eqnarray}
S_{+}^{cl}:&=&\frac{z}{1+|z|^2},\quad
S_{-}^{cl}=\frac{\bar z}{1+|z|^2},\nonumber\\
S_z^{cl}&=&\frac{1}{2}\frac{1-|z|^2}{1+|z|^2},\quad \vec
S_{cl}^2=1/4, \,(\vec S^2)_{cl}=3/4.\label{1a.3}\end{eqnarray}

There is a one-to-one correspondence between the su(2) generators
~(\ref{1a.1}) and their CS (classical) symbols given by Eqs.
~(\ref{1a.3}). Given a quantum Hamiltonian $H=H(\vec S)$, the corresponding
imaginary time phase-space action takes on the form,
\begin{equation}
{\cal A}_{su(2)}(\bar z,z)=-\,\int^{\beta}_0\langle z|
\frac{d}{dt}+H|z\rangle dt, \label{1a.4}\end{equation} with the
kinetic term being given by
$$ia=-\langle z|\frac{d}{dt}|z\rangle= \frac{1}{2}\frac{\dot{\bar z}z-\bar z\dot
z}{1+|z|^2}.$$
In particular, for the quantum $s=1/2$  Heisenberg model,
$$H=J\sum_{ij}(\vec S_i\vec S_j-1/4),$$ one gets
$$H^{cl}=\frac{J}{2}\sum_{ij}(|\langle z_i|z_j\rangle|^2-1).$$

>From the geometrical viewpoint, the su(2) coherent states
$|z\rangle$ can be thought of as sections of the magnetic monopole
bundle $P(S^2, U(1))$, with the U(1) connection one-form, $ia$,
frequently refereed to as the Berry connection. Base space of that
bundle, two-sphere $S^2$, appears as a classical phase-space of
spin, whereas its covariantly constant sections, $|z\rangle:\,
\,{\cal D}_t|z\rangle:=(\partial_t+ia)|z\rangle=0$, form a Hilbert space of a quantum spin.


\begin{thebibliography}{99}

\bibitem{sb}
S.E. Barnes, J. Phys. F {\bf 6}, 1375 (1976);
N. Read and D.M. Newns, J. Phys. C {\bf 16}, 3273 (1983);
P. Coleman, Phys. Rev. B {\bf 29}, 3035 (1984);
G. Baskaran and P.W. Anderson, Phys. Rev. B {\bf 37}, 580 (1988).

\bibitem{sf} D. Yoshioka, J. Phys. Soc. Jpn. {\bf 58}, 1516 (1989);
D.P Arovas and A. Auerbach, Phys. Rev. B {\bf 38}, 316 (1988).

\bibitem{nayak} Chetan Nayak, Phys. Rev. Lett. {\bf 85}, 178 (2000).

\bibitem{4} H. Fukuyama, Prog. Theor. Phys. Suppl. {\bf 108}, 287 (1992);
C. L. Kane et al, Phys. Rev. B {\bf 41}, 2653 (1990);
S. Feng, Z.B. Su and L. Yu, Phys. Rev. B {\bf 49}, 2368 (1994).

\bibitem{kf} E. Kochetov and A. Ferraz, Phys. Rev. B {\bf 70}, 052508 (2004).

\bibitem{optics} However, it is physically realizable in an optical lattice
\cite{zoller}.

\bibitem{zoller} D. Jaksch and P. Zoller, Ann. Phys. {\bf 315}, 52 (2005).

\bibitem{nagaoka} Y. Nagaoka, Phys. Rev. {\bf 147}, 392 (1966).

\bibitem{richmond} P.Richmond and G. Rickayzen,
J. Phys. C {\bf 2}, 528, (1969).

\bibitem{RY89} J.A. Riera and A.P. Young,
Phys. Rev. {\bf B 40}, 5285 (1989).

\bibitem{LONG-rasetti-94} M.W. Long, in "The Hubbard Model, recent results",
ed. by M. Rasetti, World Scientific, (1991).


\bibitem{36} H. Yokoyama and H. Shiba, J. Phys. Soc. Japan {\bf 56},
3670 (1987);
D. Dzierzawa and R. Fr$\acute{e}$sard, Z. Phys. {\bf B91}, 245 (1993).

\bibitem{21} E. M$\ddot{u}$ller-Hartmann,
Th. Hanisch and R. Hirsh, Physica {\bf B186-188}, 834 (1993).

\bibitem{44} E.G. Goryachev and D.V. Kuznetsov, JETP Lett. {\bf 56}, 203 (1992).

\bibitem{suto} A. Suto, Commun. Math. Phys. {\bf 140}, 43 (1991).

\bibitem{tian91} G.S. Tian, Phys. Rev. {\bf B 44} 4444 (1991);
J. Phys {\bf A 24}, 2375 (1991).

\bibitem{PLO92} W.O. Putikka, M.U. Luchini, M. Ogata,
Phys. Rev. Lett. {\bf 69}, 2288 (1992).

\bibitem{tasaki} Hal Tasaki, Progress of Theoretical Physics 99, 489 (1998);
cond mat/9712219v3.

\bibitem{becca-sorella2001}  F. Becca, S. Sorella, Phys. Rev. Lett. {\bf 86},
3396, (2001).


\bibitem{kotl} Hyowon Park, K. Haule, C.A. Marianetti, and G. Kotliar,
Phys. Rev. B {\bf 77}, 035107 (2008).

\bibitem{shastry} B.S. Shastry, H.R. Krishnamurthy, and P.W. Anderson,
Phys. Rev. B {\bf 41}, 2375 (1990).


\bibitem{vonderlinden-etal-91}
 W. von der Linden, D Edwards, J. Phys. Cond Mat. {\bf 3}, 4917 (1991),

\bibitem{basile-etal-90} A.G. Basile {\it et al}, Phys. Rev. {\bf B 41},
4842 (1990).

\bibitem{hanisch-etal-97} Th. Hanisch, G. S. Uhrig, and E. Muller-Hartmann,
Phys. Rev. {\bf B 56}, (1997).



\bibitem{scalapino} D.J. Scalapino, R. L. Sugar, S. R. White, N. E. Bickers, R. T. Scalettar,
Phys. Scr. T 27, T27, 101 (1989).

\bibitem{canada} Daniel Boies, F.A. Jackson and A.-M.S. Tremblay, Int. J. Mod. Phys. B{\bf 9}, 1001 (1995)

\bibitem{16} P. Fazekas, B. Menge, and E. M$\ddot{u}$ller-Hartmann,
Z. Phys. B {\bf 78}, 69 (1990).

\bibitem{fkm} A. Ferraz, E. Kochetov and M.Mierzejewski, Phys.
Rev. B {\bf 73}, 064516 (2006).

\bibitem{marcin}  Dr. M.Mierzejewski, private communication.

\bibitem{mattis} Daniel Mattis, Phys. Rev. {\bf 130}, 76 (1963).

\bibitem{css} This may lead to a scenario in which the FM could survive (at $U=\infty$) at any hole
concentration, provided  a Hamiltonian has a separation of charge and spin \cite{shastry}.

\bibitem{schakel} A.M.J. Schakel, Mod. Phys. Lett. B {\bf 14}, 927 (1990).

\bibitem{ranninger} M. Cuoco and J. Ranninger, Phys. Rev. B {\bf 70}, 104509 (2004).

\bibitem{zhang} X.Y. Zhang, Elihu Abrahams, and G. Kotliar,
Phys. Rev. Lett. {\bf 66}, 1236 (1991).


\end{thebibliography}
\end{document}